\documentclass[10pt,letterpaper]{article}
\usepackage[top=0.85in,left=2.75in,footskip=0.75in,marginparwidth=2in]{geometry}

\usepackage{epsfig}
\usepackage[outdir=./figures/]{epstopdf}
\graphicspath{{figures/}}
\usepackage[utf8]{inputenc}
\usepackage{amssymb}
\usepackage{amsmath}
\usepackage{booktabs}
\usepackage{array}
\usepackage{subcaption}
\usepackage{units}
\usepackage[sort&compress,square,comma,authoryear]{natbib}
\usepackage{cite}

\newcommand{\mvec}[1]{\mbox{\boldmath \ensuremath{#1}}}

\newcommand{\grace}{\textsc{grace}}
\newcommand{\gracea}{\textsc{grace-a}}
\newcommand{\graceb}{\textsc{grace-b}}
\newcommand{\kbr}{\textsc{kbr}}
\newcommand{\snr}{\textsc{snr}}

\newcommand{\mhz}{m\textsc{h}z}
\usepackage{nameref,hyperref}

\usepackage[right]{lineno}

\usepackage{microtype}
\DisableLigatures[f]{encoding = *, family = * }

\raggedright
\usepackage{changepage}

\usepackage[aboveskip=1pt,labelfont=bf,labelsep=period,singlelinecheck=off]{caption}

\makeatletter
\renewcommand{\@biblabel}[1]{\quad#1.}
\makeatother

\usepackage{lastpage,fancyhdr,graphicx}
\usepackage{epstopdf}
\fancyheadoffset[L]{2.25in}
\fancyfootoffset[L]{2.25in}

\usepackage{color}

\definecolor{Gray}{gray}{.25}

\usepackage{graphicx}
\usepackage{units}
\usepackage{textcomp}
\usepackage{booktabs}
\usepackage{sidecap}

\usepackage{wrapfig}
\usepackage[pscoord]{eso-pic}
\usepackage[fulladjust]{marginnote}
\reversemarginpar

\begin{document}
\vspace*{0.35in}

\begin{flushleft}
{\Large
\textbf\newline{Analysis of GRACE range-rate residuals with focus on  KBR instrument system noise}
}
\newline
\\
Sujata Goswami\textsuperscript{1,3*},
Balaji Devaraju\textsuperscript{1},
Matthias Weigelt\textsuperscript{1},
Torsten Mayer-G\"urr\textsuperscript{2}
\\
\bigskip
\bf{1} Institut für Erdmessung, Schneiderberg 50,
30167 Hannover \\
\bf{2} Institut of Geodesy, Technical University Graz, Austria \\
\bf{3} Max-Planck Institute of Gravitational Physics, Hannover, Germany
\\
\bigskip
* sujata.goswami@aei.mpg.de

\end{flushleft}

\section*{Abstract}
We investigate the post-fit range-rate residuals after the gravity field parameter estimation from the inter-satellite ranging data  of the \textsc{g}ravity \textsc{r}ecovery and \textsc{c}limate \textsc{e}xperiment  (\textsc{grace}) satellite mission.
Of particular interest is the high-frequency spectrum ($f >$\unit[20]{m\textsc{h}z}) which is dominated by the microwave ranging system noise. Such analysis is carried out to understand the yet unsolved discrepancy between the predicted baseline errors and the observed ones.
The analysis consists of two parts. First, we present the effects in the signal-to-noise ratio (\textsc{snr}s) of the \textsc{k-}band ranging system. The \textsc{snr}s are also affected by the \textsc{m}oon intrusions into the star cameras' field of view and magnetic torquer rod currents in addition to the effects presented by \citet{harvey2016}. Second, we analyze the range-rate residuals to study the effects of the \textsc{kbr} system noise. The range-rate residuals are dominated by the non-stationary errors in the high-frequency observations. These high-frequency errors in the range-rate residuals are found to be dependent on the temperature and effects of \textsc{s}un intrusion into the star cameras' field of view reflected in the \textsc{snr}s of the \textsc{k-}band phase observations.


\section*{Introduction}
\label{intro}
\begin{figure}[ht!]
				\centering
					\includegraphics[height = 5.2cm,%
							trim ={10 40 10 60}, clip,%
			                    		keepaspectratio]{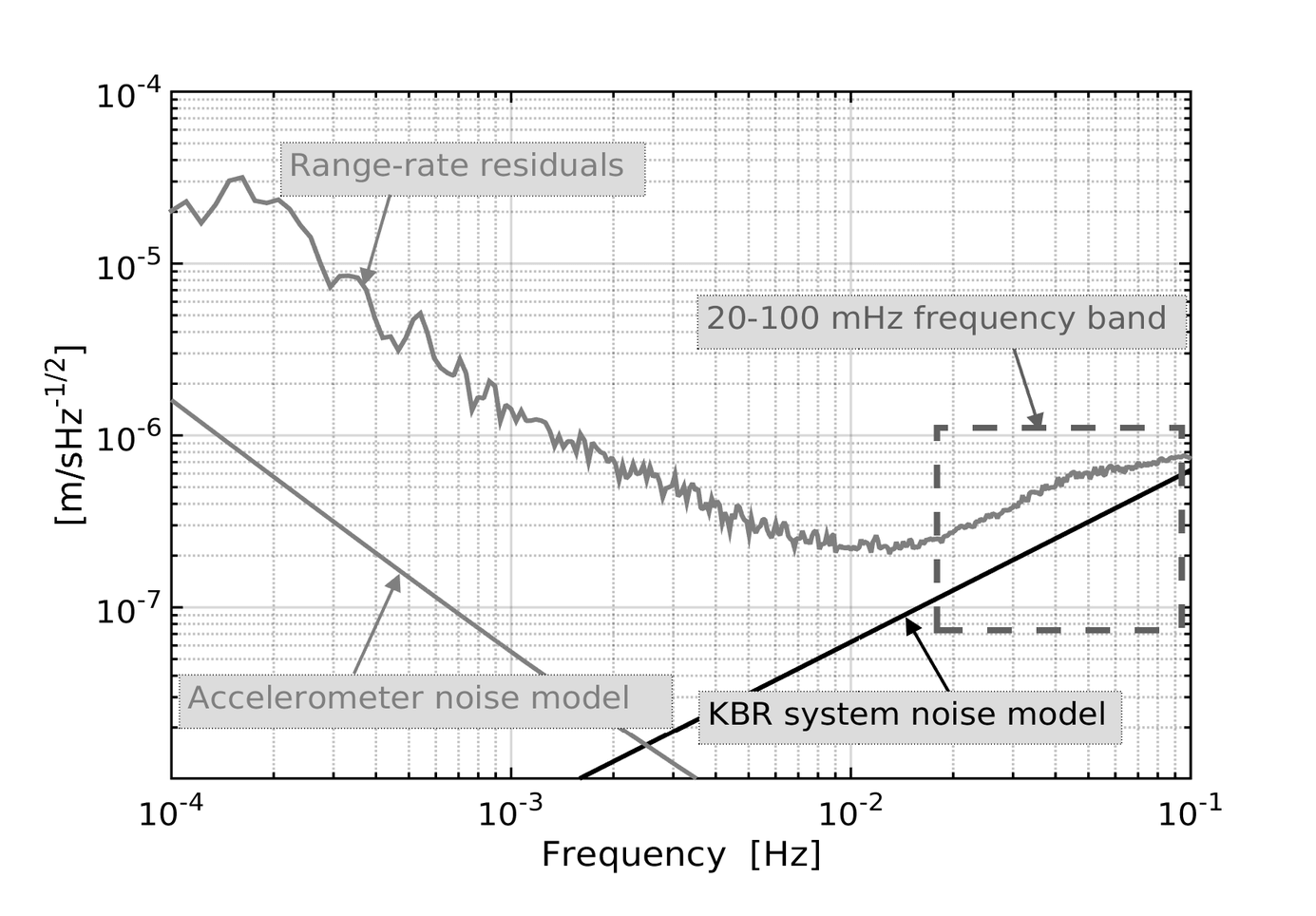}
					\caption{
    					Power spectral density of post-fit range-rate residuals plotted for December 2008 and compared with the prelaunch models of the \textsc{kbr} system noise and the accelerometer noise.}
					\label{fig:psdKBRR}
 \end{figure}

From \oldstylenums{2002}--\oldstylenums{2017}, the \textsc{grace} mission  provided  measurements of the time-variable gravity field of the \textsc{e}arth by tracking the distance between the two satellites (range) flying in a low \textsc{e}arth orbit \citep{Tapley2004}. These range observations are the main observables, which are used in the global gravity field determination. Due to their unprecedented accuracy (of a few microns)  recovery of the time-variable gravity field and the mass changes has been possible, which enabled a vast number of applications in hydrology, cryology, and climate studies \citep{ram2014,nature2013,siemes2013}. Although the accuracy of the time-variable gravity field measurements is unprecedented, still, there is an order of magnitude difference exists between the current accuracy of the \textsc{grace} solutions and the baseline accuracy that was predicted by \citet{kim2000} prior to its launch (cf. Fig. \ref{fig:psdKBRR}). Systematic errors from sensors as well as errors in the time-variable background models (cf. Table \ref{table1}) are  the primary reasons for the limited accuracy achieved in the current gravity field solutions \citep{ditmar2012,kim2000}.
Therefore, it is important to fully understand the source of these errors, which affect the accuracy of the gravity field solutions, which in turn
is required  to understand the error budget of \textsc{grace}. A full understanding of the errors in the ranging data will help  in improving the existing data pre-processing strategies, which is an important step in the global gravity field determination. Recent investigations of the star camera \citep{bandikova2014,ko2015} and accelerometer data \citep{klinger2016},  helped to improve the data  pre-processing resulted in a significant improvement in the quality of the estimated gravity field.

Pre-launch studies of the \textsc{grace} mission done by \citet{kim2000} show that the sensor noise level in the range-rate observations predominantly
consists of the accelerometer noise, star camera noise and \textsc{kbr} (\textsc{k-b}and \textsc{r}anging)
system noise. The behavior of accelerometer and \textsc{kbr} system noise was predicted in terms of their error models as shown in Fig. \ref{fig:psdKBRR}. When  the gravity field models are computed from \textsc{grace} range-rate observations, we observe the deviation between
the current error level  and the predicted error level of \textsc{kbr} system noise.
Earlier studies by \citet{1999Thomas} and \citet{phdKo2008} demonstrated that the \textsc{kbr} system noise is
dominating in the high frequencies of the range-rate observations, i.e. above
$\unit[\oldstylenums{20}]{\mhz}$. Therefore, we analyze the high-frequency range-rate observations
to study the contribution of the \textsc{kbr} system noise (highlighted in Fig. \ref{fig:psdKBRR}).

Earlier, the \textsc{kbr} system was comprehensively studied by \citet{1999Thomas} prior to the launch of the {\grace}. The performance of the \textsc{jpl} designed  \textsc{k-b}and ranging instrument had been thoroughly studied in the context of the satellite-to-satellite tracking principle.
\citet{phdKo2008} investigated the time-series of the high-frequency post-fit range-rate residuals and provided initial strategies for analyzing sensor noise.
This was followed by an analysis of the \textsc{s}ignal-to-\textsc{n}oise \textsc{r}atio (\snr) of the ranging system \citep{ko2012}, which correlated the poor {\snr} values of the ranging system with the high-frequency range-rate residuals.
However, the study did not establish the source of the poor {\snr} values.
We investigated the source of the {\snr} variations and attributed them to the sun intrusions into the star camera and temperature variations of the accelerometer \citep{goswami}. We found that the \textsc{snr} variations in the \textsc{k-}band frequency of \textsc{grace-b} due to temperature effects degrade the quality of ranging observations, which is reflected in the range-rate residuals.
\citet{harvey2016} analyzed the {\snr} data from \oldstylenums{2006}--\oldstylenums{2013} and they identified that the variations in the temperature, measured by one of the thermistors located near the ranging system, were affecting the \textsc {\snr} of the \textsc{k-}band frequency of  \textsc{grace-b} (see plot \textsc{k-b} in Fig. \ref{fig:2k7Kflag_A}(b.)). They also showed an impact of the sun intrusions into the {\snr}s of the \textsc{k-}band frequencies of \textsc{grace-b} and \textsc{k}a-band frequency of \textsc{grace-a} (cf. \textsc {\snr} plots (\textsc{k}a-\textsc{a}, \textsc{k-a}, \textsc{k-b}) in Fig. \ref{fig:2k7Kflag_A}). Since the \textsc{k}a-band \textsc{snr} values of \textsc{grace-b} were anomalous during the analyzed time-period (cf. plot \textsc{k}a-\textsc{b} in Fig. \ref{fig:2k7Kflag_A}(b.)),  no characteristics were analyzed.
The study defined the characteristics of the \textsc{snr}s mainly in the context of the mission requirements.
Earlier,  \citet{ditmar2012} studied the noise in the \textsc{grace} sensor data by analyzing the \textsc{p}ower \textsc{s}pectral \textsc{d}ensity (\textsc{psd}) of the range-rate residuals.  The error budget was presented for year \oldstylenums{2006} using the noise models based on the \textsc{psd}s of the range-rate residuals.  \citet{inacio} analyzed the \textsc{grace} star camera errors from year \oldstylenums{2003} to \oldstylenums{2010} and presented an approximate budget  of the star camera errors in the gravity field solutions.

In this study, our approach is to analyze the post-fit range-rate residuals, in particular the residuals in the frequencies above \unit[\oldstylenums{20}]{m\textsc{h}z}, after the gravity field parameter estimation from the real \textsc{grace} data. By analyzing the post-fit range-rate residuals, we aim to understand the sources of the noise in the range-rate observations, as they reflect the errors present in the \textsc{grace} data, at least partially \citep{goswami2016a}.
With this approach of analysis, we show the characteristics that were not seen in the earlier studies based on  the \textsc{psd} analysis of the \textsc{grace} data. Specifically, we analyze the range-rate residuals and the required \textsc{grace} data in the \emph{argument of latitude} and time domain. The \emph{argument of latitude} is defined as the angle between ascending node and the satellite at an epoch. For more details, please refer to \citet{orbits}.
Plotting the satellite observations along the \emph{argument of latitude} and time helps us to analyze their systematic behavior.
An example of such a plot is shown in Fig. \ref{fig:2k7Kflag_A}  where observations are plotted along the \emph{argument of latitude} on the vertical axis  varying from \unit[\oldstylenums{0}-\oldstylenums{360}]{degrees} (bottom `\textsc{ae}' to top `\textsc{ae}' where `\textsc{ae/de}' is ascending equator/descending equator and `\textsc{np/sp}' is north pole/south pole) and time in days on the horizontal axis.  We present results of the \textsc{grace} observations of year  \oldstylenums{2007} and \oldstylenums{2008}, i.e. two years.
The solar flux was low during that period, which minimizes its impact on the data and is hence a good candidate for non-stationary error analysis \citep{uli2016}.

 Our contribution focuses on the following points
\begin{enumerate}
\item  Unidentified effects in the \textsc{snr}s.
\item  The analysis of post-fit range-rate residuals with focus on the non-stationary errors in the high frequencies and their contribution to the parameters estimated during the gravity field parameter estimation process.
\end{enumerate}
Our main contribution is the analysis of the post-fit range-rate residuals with focus on the \textsc{kbr} system noise. In order to understand the  \textsc{kbr} system noise it is important to understand the \textsc{snr}s of the four frequencies of the \textsc{kbr} microwave ranging system. Therefore, we analyzed the \textsc{snr}s and found that there  are effects in \textsc{snr} related to the \textsc{m}oon intrusions and magnetic-torquer rod currents. These effects are in addition to those previously identified by \citet{harvey2016}.
 First, we present those effects in the \textsc{snr} values and  then we present an analysis of the high-frequency spectrum of post-fit range-rate residuals where the \textsc{k-}band ranging system noise is dominating.
The outline of our contribution is as follows.  We discuss  the \textsc{m}oon intrusions and magnetic-torquer rod currents in Section \ref{section1}.
Further, we discuss our gravity field parameter estimation scheme in Section \ref{section2} followed by an analysis of the post-fit range-rate residuals with focus on the high-frequency errors in Section \ref{noise}. The contribution of the high-frequency errors in the estimated parameters is discussed in Section \ref{prefit}.

\section{Unidentified effects in the SNRs \label{section1}}
The range observations are computed by the combination of   \textsc{k-} and \textsc{k}a-band phase observations from  the two satellites. In order to investigate the system noise of the \textsc{grace}  \textsc{k-}band ranging system, it is important to understand the quality of  these four phase observations.  The \textsc{snr} values of these observations reflect the signal strength and ranging measurement quality. They are also used to filter the spurious phase measurements when combining the four phase measurements to get the inter-satellite range data. The combination is performed by \textsc{jpl}  during the \textsc{kbr} level \oldstylenums{1}\textsc{a} to level \oldstylenums{1}\textsc{b} processing. The \textsc{snr} values are expressed as a factor of \unit[]{\oldstylenums{0.1}db-\textsc{h}z} in the standard \textsc{kbr} level   \oldstylenums{1}\textsc{b} data.
The phase measurements below \textsc{snr} values \unit[\oldstylenums{340}]{\oldstylenums{0.1}db-\textsc{h}z} are  considered as spurious and  are therefore  not considered in the combination of phase measurements.  This means either we see a gap in the data or interpolated values depending upon the length of the time interval  \citep{atbd-grace}. \newline
According to  \citet{harvey2016},   the \textsc{snr} value is defined as the amount of power in  \unit[\oldstylenums{20}]{ms} integrations of signal power (integrated against a phase locked local model) compared to an integration with the local model in quadrature. The  minimum \textsc{snr} requirement for the \textsc{grace} mission is \unit[\oldstylenums{63}]{db-\textsc{h}z} or \unit[\oldstylenums{630}]{\oldstylenums{0.1}db-\textsc{h}z} (as given in the standard \textsc{kbr} level \oldstylenums{1}\textsc{b} data).

The \oldstylenums{1}-$\sigma$ (\oldstylenums{1}-phase) error corresponding to the \textsc{snr} for \unit[\oldstylenums{1}]{s} time interval data is given as
		\begin{equation}
				\sigma^{\text{K/Ka}}_{\phi,i} = \frac{1}{2 \pi 		(\textsc{snr}^{\text{K/Ka}}_{i})}
				\label{eqn:1phaseKa}
		\end{equation}
 in units of cycles \citep{1999Thomas} where `$i$' $\in \text{\{A, B\}}$.
Eqn. \ref{eqn:1phaseKa} demonstrates that low \textsc{snr}s can lead to high noise in the phase observations. The total noise of the phase observations constitutes the \textsc{k-}band ranging system noise (see \citet{1999Thomas} for details), which dominates in high frequencies (above \unit[\oldstylenums{20}]{m\textsc{h}z}) of the range-rate observations.
In order to understand the error characteristics of the range-rate residuals in high frequencies, we therefore need to analyze the \textsc{snr}s of the phase observations. In the following subsections, we present the  systematic effects  that are not yet discussed in the existing literature.
	\subsection{Moon intrusion effects on SNR \label{moon}}
\begin{figure}[htbp!]
      		\centering
      			\includegraphics[width=0.85\linewidth, keepaspectratio]{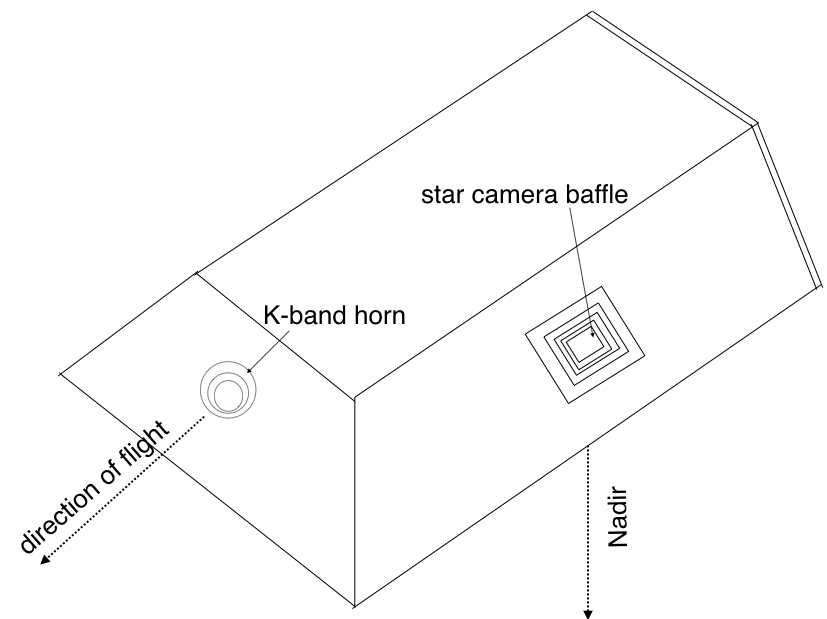}
			\caption{
      				Star camera baffle structure on the \textsc{grace} spacecraft after  \citet{harvey2016}.
      				} \label{fig:baffle}
		\end{figure}

The \textsc{grace} star cameras are blinded by  the \textsc{s}un and the \textsc{m}oon  every \unit[\oldstylenums{161}]{d} and \unit[\oldstylenums{27}]{d} respectively \citep{TamaraPhD}. These effects are called \textsc{s}un and \textsc{m}oon intrusions. Each spacecraft has two star cameras on board, designated as \emph{head\#1} and \emph{head\#2}, which are located on the lateral side of each spacecraft. In Fig. \ref{fig:baffle}, the star camera baffle represents the location of one of the star camera heads on that lateral side.
During the in-flight attitude control, when one of the star camera heads is blinded, the other head is set as primary star camera, which is available for the attitude determination. When both star camera heads are available, the attitude of the spacecraft is obtained by combining the data of the two star camera heads which is done during the ground processing \citep{TamaraPhD,romans2003}.
	\begin{figure}[htbp!]
      		 (a.) \\	\includegraphics[width=0.85\linewidth, keepaspectratio]
      				  						{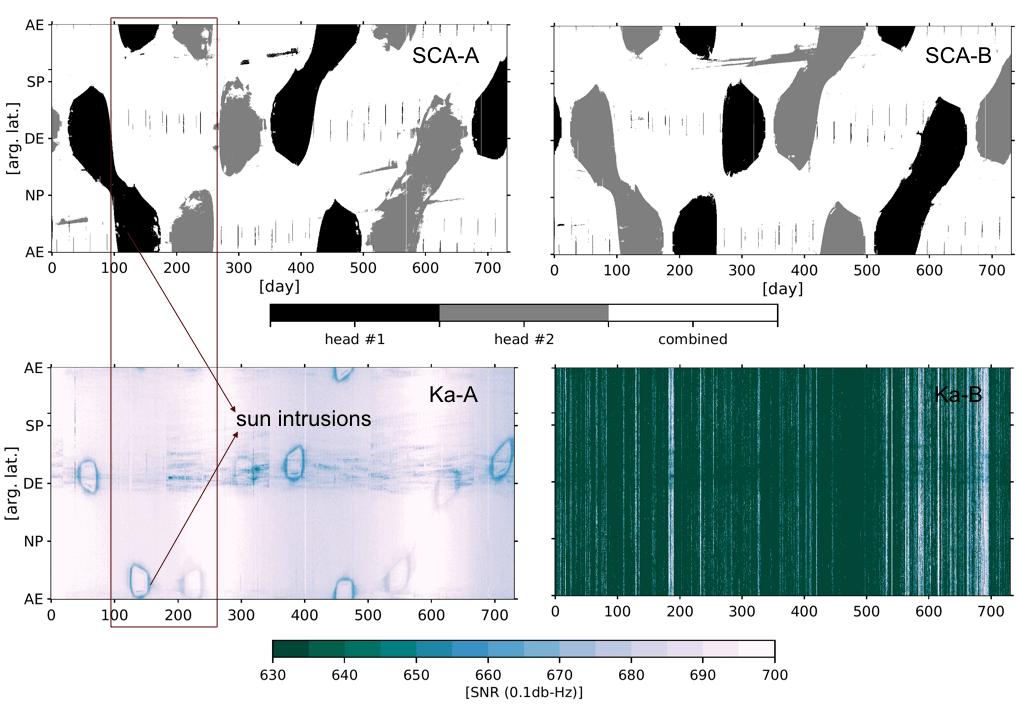} \\
 			     (b.) \\    \includegraphics[width=0.85\linewidth, keepaspectratio]
     			  						{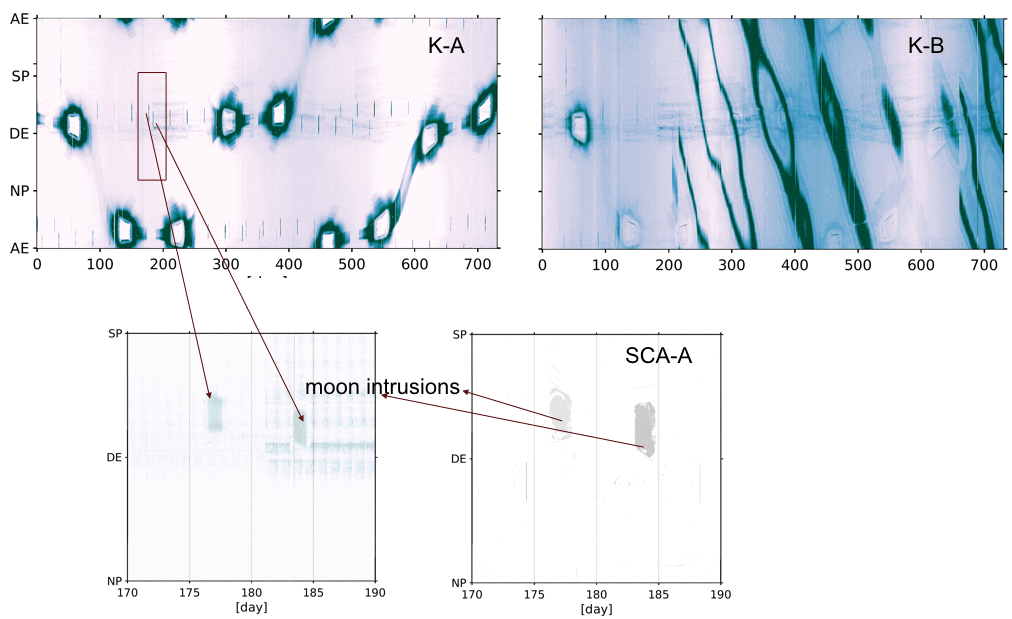}
      			\caption{
                		The top row shows when the star camera heads were blinded during the year 2007 and 2008, where \emph{gray} color and  \emph{black} color represent the blinding of the \emph{head\#2} and \emph{head\#1} respectively and \emph{white} means both star camera heads are available for the attitude determination.
                    The \textsc{k}a-band \textsc{snr}s of \textsc{grace-a} and \textsc{b} are shown in second row and the \textsc{k-}band \textsc{snr}s are shown in third row. The fourth row shows a zoomed-in view of the \textsc{m}oon intrusions into \textsc{snr} and blinded heads of star camera of corresponding spacecraft.
      				    } \label{fig:2k7Kflag_A}
		\end{figure}

 Fig. \ref{fig:2k7Kflag_A}a (\textsc{sca-a} and \textsc{sca-b}) shows the \textsc{s}un and \textsc{m}oon intrusions into the star camera of each satellite during the year \oldstylenums{2007} and \oldstylenums{2008}. Black color represents the periods when \emph{head\#1} was blinded and \emph{head\#2} was active, gray color represents the periods when \emph{head\#2} was blinded and \emph{head\#1} was active, white color represents the periods when none of the heads were blinded.

As shown in the same figure (cf. Fig. \ref{fig:2k7Kflag_A}), all the three valid \textsc{snr}s (\textsc{k-b, k-a}, \textsc{k}a-\textsc{a}) of both spacecraft experience a drop in their values during the intrusions into the star camera. As previously mentioned, the \textsc{k}a-band \textsc{snr}  of \textsc{grace-b} (\textsc{k}a-\textsc{b}) was anomalous during this time and, thus, we do not observe any related characteristics in these values. Here we focus only on the affect of the \textsc{m}oon intrusions on the \textsc{snr}s, which are highlighted in the \textsc{k-}band  \textsc{snr} of \textsc{grace-a} (Fig. \ref{fig:2k7Kflag_A}, bottom left panel) and the corresponding star camera data flags plotted for the same duration (Fig. \ref{fig:2k7Kflag_A}, bottom right panel). The \textsc{k}a-band \textsc{snr} of \textsc{grace-a} and  \textsc{k}-band \textsc{snr} of \textsc{grace-b} also suffered from \textsc{m}oon intrusions, but their values did not drop below mission requirements (\unit[\oldstylenums{630}]{\oldstylenums{0.1}db-\textsc{h}z}). The \textsc{snr}s were still ranging between \unit[\oldstylenums{675}-\oldstylenums{685}]{\oldstylenums{0.1}db-\textsc{h}z}. However, the  \textsc{k}-band \textsc{snr} of \textsc{grace-a} dropped much lower (ranging between \unit[\oldstylenums{630}-\oldstylenums{640}]{\oldstylenums{0.1}db-\textsc{h}z}) than the other two \textsc{snr}s during \textsc{m}oon intrusions, sometimes even below mission requirements (for example, \textsc{m}oon intrusions between the days \oldstylenums{400}-\oldstylenums{500}).

Ring shape structure during the \textsc{m}oon intrusions in the \textsc{snr}s which we can see in the zoomed-in plot in the bottom panel of Fig. \ref{fig:2k7Kflag_A}(b), are similar to the ring shape structures during the \textsc{s}un intrusions in the \textsc{snr}s. In the bottom panel of Fig. \ref{fig:2k7Kflag_A}(b), \textsc{m}oon intrusions are shown in \textsc{snr} during days from \oldstylenums{170} to \oldstylenums{190}.  The signatures of \textsc{m}oon intrusions resemble the physical structure of the star camera baffle (cf. Fig. \ref{fig:baffle}), similar to the signatures left by \textsc{s}un intrusions.

Since the \textsc{s}un intrusions effects were identified in \textsc{grace} before the launch of \textsc{grace-f}ollow \textsc{o}n (\textsc{grace-fo}), the \textsc{grace-fo} \textsc{kbr} assembly will be shielded to protect it from the interference caused by the \textsc{i}nstrument \textsc{p}rocessing \textsc{u}nit (\textsc{ipu}). Therefore, we do not expect to find the \textsc{m}oon intrusion effects on \textsc{snr} in  \textsc{grace-fo} data (personal communication, Gerhard L. Kruizinga, \textsc{jpl, nasa} on \oldstylenums{10} \textsc{o}ct. \oldstylenums{2016}).
	\subsection{Magnetic torquer rod current effects on SNR \label{torque}}
Three magnetic torquer rods (\textsc{mtq}) located off-center in each \textsc{grace} spacecraft, mounted parallel to the satellite body reference triad, serve as the primary attitude control actuators. Magnetic torquers generate a magnetic dipole \textbf{m}, whose magnitude is dependent on the applied electric current. The resulting torque \textbf{T} acting on the spacecraft is then given as the vector product of the sum of the  magnetic dipoles generated by all three \textsc{mtq}s  and the \textsc{e}arth\textquotesingle s magnetic flux density \textbf{B} (cf. Eqn. \ref{torq}) \citep{wertz}.
\begin{equation}
\textbf{T} = \textbf{m} \times \textbf{B}
\label{torq}
\end{equation}

\begin{figure}[htbp!]
			\centering
    		\includegraphics[width=1\linewidth, keepaspectratio]
        			{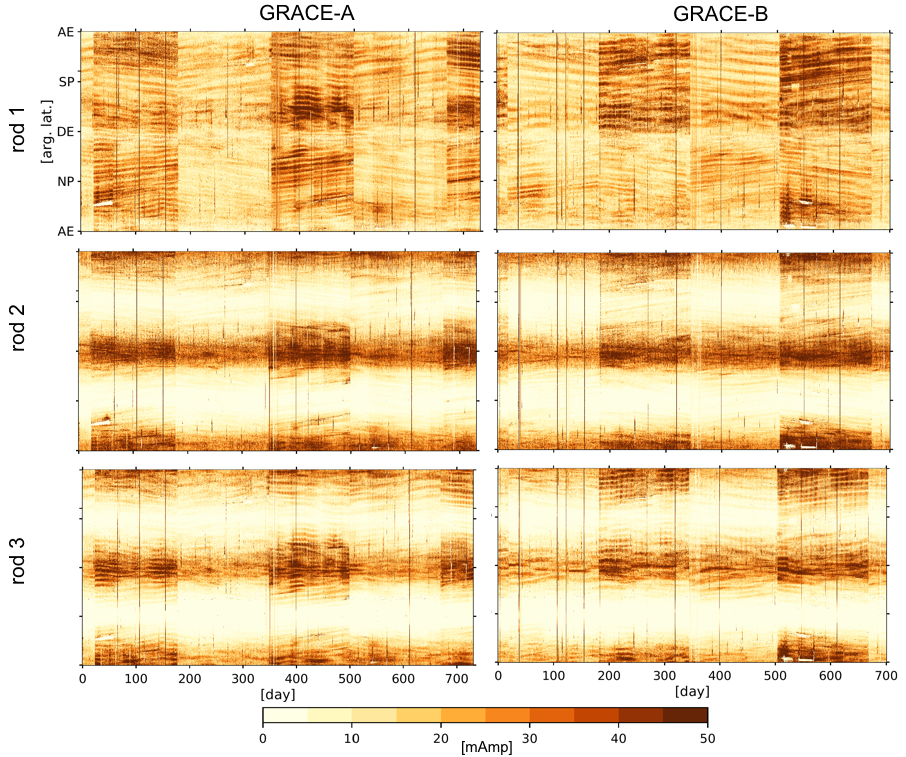}
			\caption{
        			Absolute value of currents in the three magnetic torquer rods of \textsc{grace-a} and \textsc{grace-b} for 2007 and 2008. Periods of 161 days with high currents along the equator (descending and ascending) can also be seen which are related to satellite attitude.
        			} \label{fig:2k78Ktor_AB}
		\end{figure}

In Fig. \ref{fig:2k78Ktor_AB}, the absolute value of the magnetic torquer rod currents of both satellites are plotted along the argument of latitude for the years \oldstylenums{2007} and \oldstylenums{2008}.  The variation in the currents every \unit[\oldstylenums{161}]{d} is dependent on the primary star camera head during attitude determination and the accuracy of the attitude observed by it.

According to \citet{herman2004, TamaraPhD}, during  \oldstylenums{2007} and  \oldstylenums{2008} attitude observed by \emph{head\#2} was more accurate than that of the attitude observed by \emph{head\#1} of both spacecraft.
When \emph{head\#2} was used in the attitude control loop,  less torque was needed to keep the satellite attitude within the limits required for inter-satellite pointing. Therefore,  the electric currents flowing through the \textsc{mtq}s were smaller during the period when \emph{head\#2} was available. During
the period when \emph{head\#1} was used in the attitude control loop, more electric currents were needed. As a result of the differences between the accuracies of the two star camera heads on board each spacecraft, we see the alternate \unit[\oldstylenums{161}]{d} period variations in the magnitude of the electric currents flowing through the three rods of each spacecraft (cf. Fig. \ref{fig:2k78Ktor_AB}).
\begin{figure}[h!]
		\centering
		\includegraphics[width=0.99\linewidth, keepaspectratio]{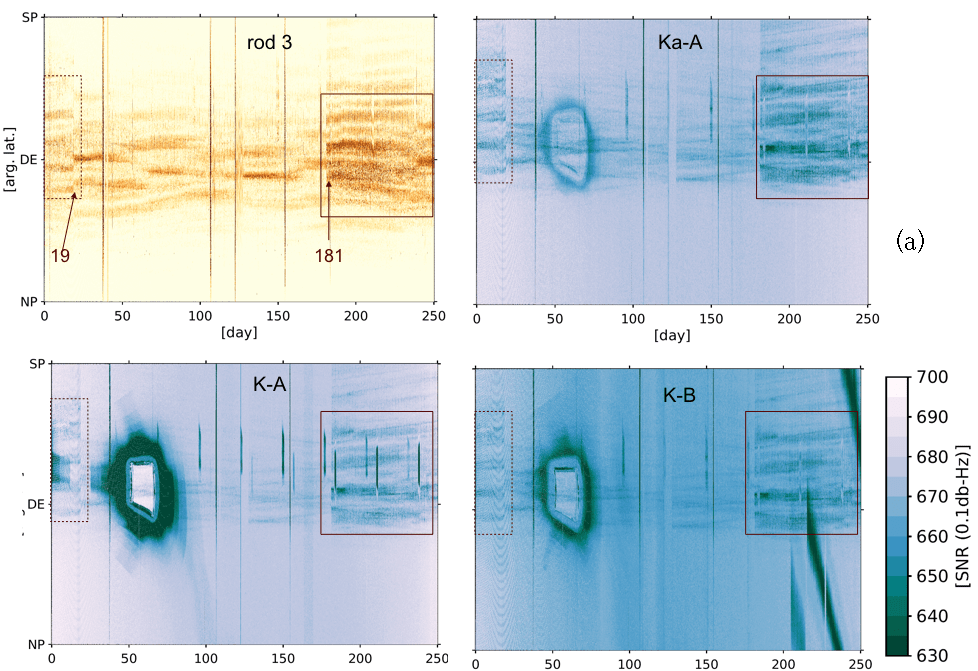}
  		\begin{minipage}[c]{0.36\linewidth}
  		\includegraphics[width= 0.99\linewidth]{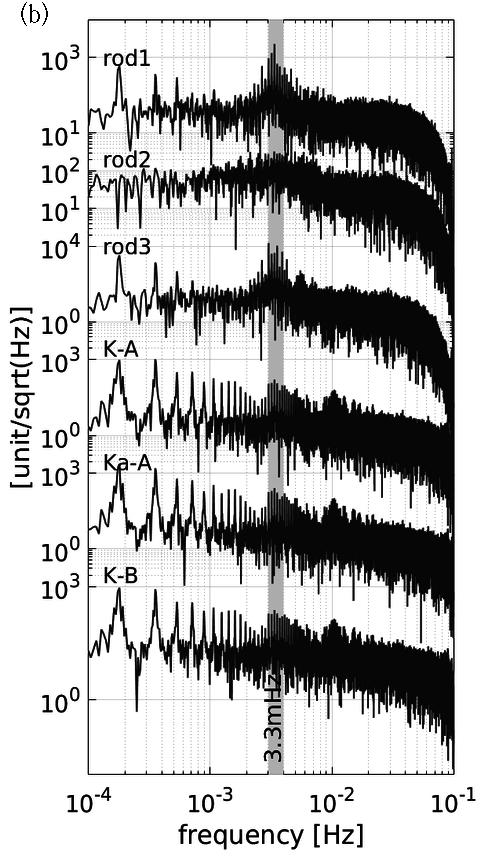}
  		\end{minipage}\hfill
  		\begin{minipage}[c]{0.62\linewidth}
  	  	\caption{%
      		    (a) \emph{Above}: %
      				Effects of the currents of rod 3 of \textsc{{\graceb}} on the three \textsc{snr}s of the \textsc{kbr} microwave system. The effects are shown for 250 days of year 2007 highlighted with the regions with high torquer rod currents and the affected \textsc{snr}s due to them. The color scale of the plotted currents of the rod 3 are same as that of the currents plotted in Fig. \ref{fig:2k78Ktor_AB}. %
            			\newline%
            				(b) \emph{Left}: %
  	    				The power spectral densities of the three magnetic torquer rod currents of \textsc{{\graceb}} and the three \textsc{snr}s (\textsc{k}- and \textsc{k}a-band \textsc{snr} of \textsc{{\gracea}} and \textsc{k}-band \textsc{snr} of \textsc{{\graceb}}). All the three \textsc{snr}s also show a peak at the frequency \unit[3.3]{m\textsc{h}z} which is the dominant frequency of the currents flowing through \textsc{mtq}s.
  	  		} \label{fig:psd_torsnr}
  	\end{minipage}
	\end{figure}

The three valid \textsc{snr}s, which are  \textsc{k}a-band \textsc{snr} of \textsc{grace-a}, \textsc{k-}band \textsc{snr} of \textsc{grace-a} and \textsc{b}, are observed to be affected by the \textsc{mtq}s of \textsc{grace-b}. The three valid \textsc{snr}s of both spacecrafts are found to be correlated with the currents flowing through rod \oldstylenums{2} and \oldstylenums{3} of \textsc{grace-b}. In Fig. \ref{fig:psd_torsnr}(a) we show the correlations present between rod \oldstylenums{3} of  \textsc{grace-b} and the three valid  \textsc{snr}s of both spacecrafts as a zoomed-in picture for the \unit[\oldstylenums{250}]{days} of year \oldstylenums{2007}. The currents were  smaller between days \oldstylenums{19} to \oldstylenums{180} as opposed to days from \oldstylenums{181} to \oldstylenums{250}. This is because the primary star camera head from   day   \oldstylenums{19} to day \oldstylenums{180} was \emph{head\#2} and beyond that it was  \emph{head\#1} on \textsc{grace-b}  (see Fig. \ref{fig:2k7Kflag_A} for the details of primary star camera heads). During the period of strong currents (from day \oldstylenums{181} to \oldstylenums{250}) flowing through rod \oldstylenums{3} of \textsc{grace-b}, their effect on the \textsc{snr}s can be seen easily in all the three \textsc{snr} plots as opposed to the period when small currents were flowing (from the day \oldstylenums{19} to \oldstylenums{180}) through the torquer rods (see highlighted region in Fig. \ref{fig:psd_torsnr}(a)). Here, we see that the currents flowing through the \textsc{mtq}s have an impact on the  \textsc{snr}s, which is clearly visible in the \textsc{snr}s in Fig. \ref{fig:psd_torsnr}. However, there is no drop observed in the \textsc{snr}s below  mission requirements (\unit[\oldstylenums{630}]{\oldstylenums{0.1}db-\textsc{h}z}) during any of the alternate \unit[\oldstylenums{161}]{d} cycle of currents in \textsc{mtq}s.

The power spectral densities of the three valid \textsc{snr}s show large values around the frequency \unit[\oldstylenums{3.3}]{\mhz} which was already found to be associated with the magnetic torquer rod currents of the \textsc{grace} spacecrafts by \citet{pointing-grace}.

\begin{figure}
\includegraphics[width=0.9\linewidth, keepaspectratio]{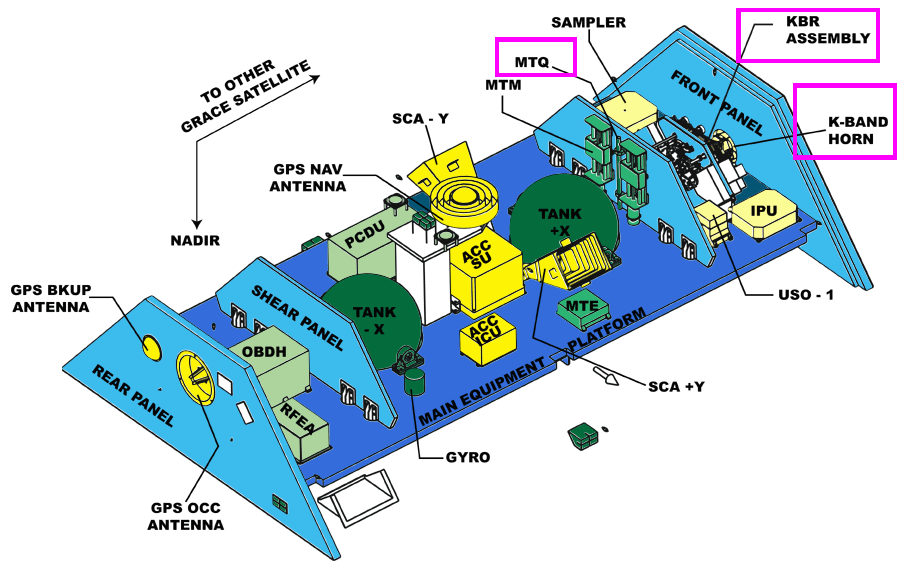}
\caption{Internal view of \textsc{grace} showing the location of the \textsc{mtq}s near the front panel where \textsc{kbr} assembly is mounted. \textcircled{c} \url{https://directory.eoportal.org/web/eoportal/satellite-missions/g/grace}}
\label{grace}
\end{figure}

In  Fig. \ref{grace}, we see that the \textsc{kbr} assembly is located near one of the \textsc{mtq} rods.  It is possible that the currents flowing through the rod are causing the electromagnetic interference that affects the \textsc{kbr} assembly. Thus, we see a correlation between the \textsc{mtq} current and variations in the \textsc{snr}s. However, this hypothesis has to be studied further. Investigations related to \textsc{mtq} rod current effects on primary sensors (accelerometer, star-trackers, \textsc{kbr} assembly) are ongoing in \textsc{jpl, nasa} (personal communication, Gerhard L. Kruizinga, \oldstylenums{10} \textsc{o}ct. \oldstylenums{2017}).
\section{Analysis of the post-fit range-rate residuals \label{section2}}
In this section, we discuss the errors absorbed by the high-frequency range-rate residuals $(f > \unit[\oldstylenums{20}]{\mhz})$ and their possible sources. Here, we analyze the range-rate residuals that are obtained after the full parameter estimation chain of the gravity field parameter estimation.

\noindent The gravity field parameters are estimated using the standard \textsc{itsg-\oldstylenums{2014}} processing chain \citep{itsg2014}. The unconstrained monthly solutions are estimated up to degree \oldstylenums{60} using the  variational equations approach. For details  regarding the implementation of the approach see \citet{torstenphd}. The unknown parameters are estimated using least-squares variance component estimation \citep{koch2002}. These unknown parameters include the Stokes's coefficients, initial orbital state parameters, accelerometer scale and bias parameters. The orbital state (\mvec{r}, \mvec{\dot r}) and the accelerometer parameters are estimated once per day along the three axes (x, y, z).

The basic least-squares adjustment is as follows --
	\begin{align}
		\mvec{\Delta{{l}}} 		  &=  \mvec{\Delta{{\hat{l}}}} + \mvec{\hat{e}}_{\text{\textsc{kbr}}} + \mvec{\hat{e}}_{\text{\textsc{gps}}} + \mvec{\hat{e}}_{\text{\textsc{acc}}},   \label{eq2} \\
		\mvec{\Delta{{\hat{l}}}}  &=  \mathbf{A}\, \mvec{\Delta{\hat{x}}}
		\label{eq:parest}
	\end{align}

          where, $\mathbf{A}$ is the design matrix of size (i$\times$j), (i, j)$\in$(rows, columns).
          \newline $\mvec{\Delta{\hat{x}}}$ consists of estimated Stokes's coefficients ($\text{c}_{nm}$, $\text{s}_{nm}$), accelerometer scale and bias parameters, and orbital state parameters (\mvec{r}, \mvec{\dot r}).
          \newline $\mvec{\Delta{{l}}}$ are the  reduced range-rate observations $(\delta{\dot{\rho}})$, \textsc{gps} observations containing satellite state parameters (\mvec{r}, \mvec{\dot r}) and accelerometer scale and bias parameters.
          \newline $\mvec{\hat{e}}_{\text{\textsc{kbr}}}, \mvec{\hat{e}}_{\text{\textsc{gps}}}, \mvec{\hat{e}}_{\text{\textsc{acc}}}$ are the range-rate residuals, orbital state residuals, accelerometer scale and bias residuals respectively.

Reduced range-rate observations $(\delta{\dot{\rho}})$ used in parameter estimation are computed as -
\begin{align}
		 \delta{\dot{\rho}}  &= \dot{\rho} - \dot{\rho}_{0}
\end{align}
where, $\dot{\rho}$ and  $\dot{\rho}_{0}$ are the observations obtained from the satellite and  observations computed from the dynamic orbit that is obtained from the  state-of-the art background models, respectively. Background models used to compute $ \dot{\rho}_{0}$ are mentioned in Table \ref{table1}.
We use the term \emph{`pre-fit range-rate residuals'}  for reduced  range-rate observations $(\delta{\dot{\rho}})$ and \emph{`post-fit range-rate residuals'} for the range-rate observations obtained as residuals of the reduced range-rate observations after least-squares parameter estimation fit denoted as ($\mvec{\hat{e}}_{\textsc{\text{kbr}}}$) in Eqn. \ref{eq:parest}. In the following sections, we use the notation ($\mvec{\hat{e}}$)  to refer to the post-fit residuals of range-rate observations ($\mvec{\hat{e}}_{\textsc{\text{kbr}}}$).
 \begin{table} [htbp!]
  	\caption{
    	Background models (perturbations)  that are reduced from the range-rate observations during gravity field processing.
    		}
  	\begin{center}
   		\begin{tabular}{l l}
   		\toprule
   		\textbf{Models} & \textbf{Standards} \\ [0.5ex]
   		\midrule
   		Earth rotation & IERS 2010 \citep{iers2010}  \\
   		Moon, \textsc{s}un and planet ephemeris & JPL DE421  \citep{de421}\\
	    	Earth tide &  IERS 2010 \citep{iers2010} \\
		Ocean tide & EOT11a  \citep{eot11a} \\
		Pole tide & IERS 2010 \citep{iers2010} \\
		Ocean pole tide  & Desai 2003 \citep{iers2010}  \\
		Atmospheric tides (S1, S2)  & Bode-Biancale 2003 \citep{s1s2} \\
		Atmosphere and Ocean Dealiasing  &  AOD1B RL05  \citep{aod1b}\\
		Relativistic corrections & IERS 2010 \citep{iers2010}\\
		Permanent tidal deformation & includes (zero tide) \\ [1ex]
   		\bottomrule
    	\end{tabular} \label{table1}
  	\end{center}
 \end{table}
	\begin{figure}[htbp!]
		\centering
		\includegraphics[width = 0.80\linewidth, keepaspectratio]
				{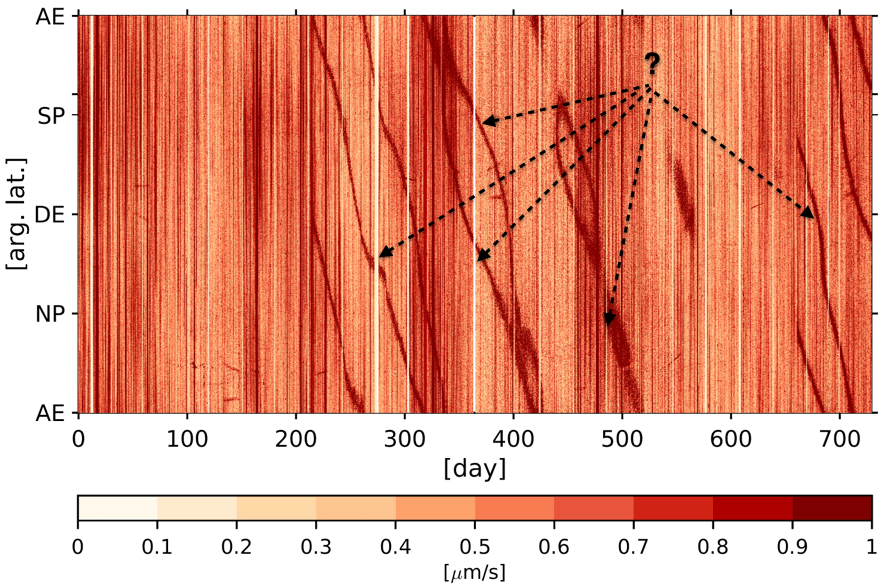}
		\caption{
			Post-fit range-rate residuals computed using the \textsc{itsg-\oldstylenums{2014}} parameter estimation chain and are plotted on an absolute scale. The residuals are plotted for the two year duration starting from \oldstylenums{1} \textsc{j}anuary \oldstylenums{2007}.
			} \label{fig:postf}
	\end{figure}
	\subsection{Error characteristics of the high-frequency range-rate residuals \label{noise}}
	This section focuses on understanding of the error characteristics of  high-frequency post-fit range-rate residuals ($>$\unit[\oldstylenums{20}]{\mhz}) and identifying their sources.
	As seen in Fig. \ref{fig:postf}, one of the most interesting features in the post-fit range-rate residuals is the pattern of bands with high value of post-fit residuals, which begins from day \oldstylenums{200} and continues until the end of \textsc{d}ecember \oldstylenums{2008} (day \oldstylenums{730}). The structure of these bands changes and repeats after a shift in time.

    Here, we are interested in understanding:
    \begin{itemize}
    \item[--] why is the amplitude of residuals high in certain regions which vary over the orbit and time?
    \item[--] and in which frequencies do these errors lie? It is important to know whether they are affecting the most important frequency band of the large time-variable gravity field signal, i.e., \unit[\oldstylenums{0.1}-\oldstylenums{18}]{\mhz} \citep{1999Thomas}.
    \end{itemize}

\begin{figure}[htbp!]
		\centering
		\includegraphics[width = 0.99\linewidth, keepaspectratio]
				{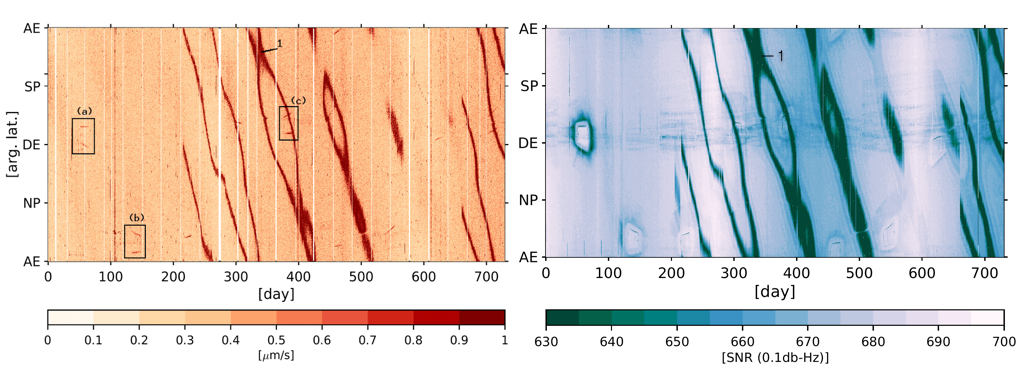}  \\
		\includegraphics[width = 0.90\linewidth, keepaspectratio]
				{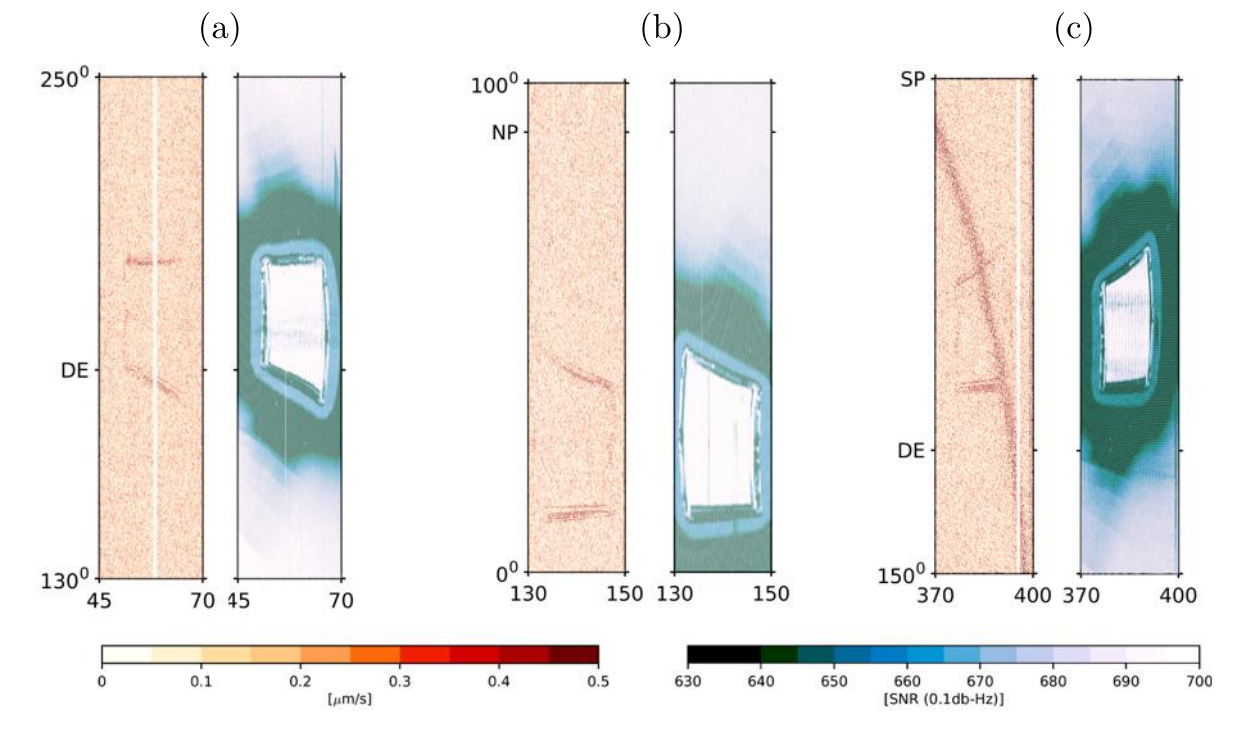}
		\caption{
			\emph{Above}: (\emph{right}) Absolute of the high-frequency post-fit residuals ($\mvec{\hat{e}_{\text{HP}}}$) plotted for year 2007 and 2008 along the argument of latitude and time in days along with the \textsc{k-}band \textsc{snr} of \textsc{grace-b} (\emph{left}); temperature dependent bands are marked as `1' and `(a)', `(b)', `(c)' are the effects related to the \textsc{s}un intrusions; \emph{below}: are the zoomed-in picture of the residuals correlated with the \textsc{s}un intrusions related effects in the \textsc{snr}s. The \textsc{k}-band \textsc{snr} of \textsc{grace-a} is plotted here to show the correlation with postfits.}
			 \label{fig:intruPfits}
	\end{figure}

Investigation of the post-fits revealed that these features are dominating in the frequencies above \unit[\oldstylenums{20}]{\mhz} which are plotted in Fig. \ref{fig:intruPfits}. The filters applied on the post-fit range-rate residuals are provided by  \citet{ltpda}.

\begin{figure}[htbp!]
		\centering
			\includegraphics[width=1.05\linewidth, keepaspectratio]{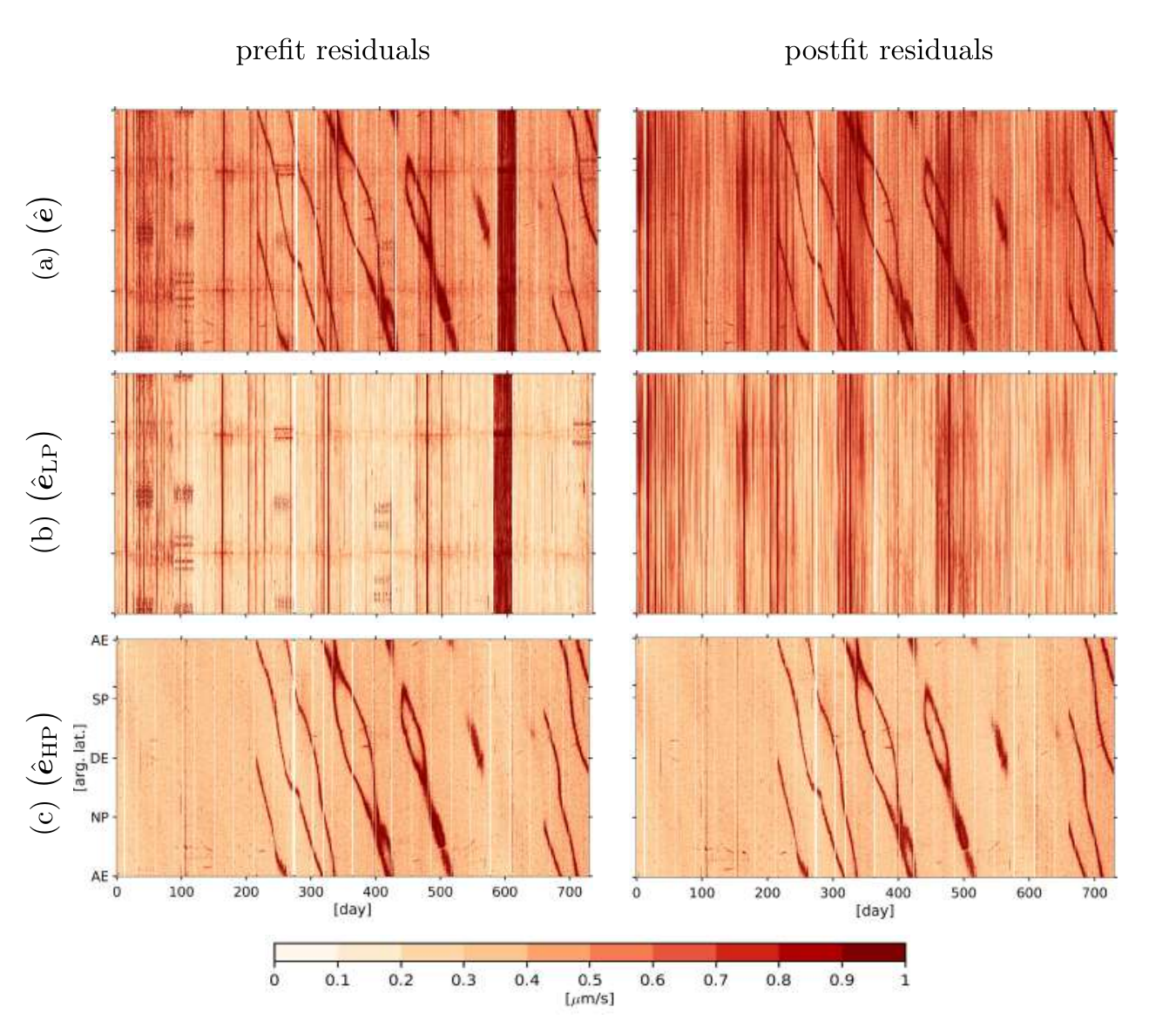}
		\caption{
			Comparison of the absolute values of pre-fit and post-fit residuals (a) and their low-pass (b) and high-pass filtered parts (c). The darkest patch from the day 578 to 609 shows that the prefits were comparatively higher for the month of August 2008 than the other months. However, the solution converged with the noise level comparable to other months as can be seen in the postfits.
			} \label{fig:fullspectra}
		\end{figure}

We denote the set of high-pass filtered post-fit range-rate residuals as $(\hat{e}_{\text{HP}})$ and low-pass filtered post-fit range-rate residuals as $(\hat{e}_{\text{LP}})$. The set of low-pass filtered $(<\unit[\oldstylenums{20}]{\mhz})$ post-fit range-rate residuals  does not contain these features. Comparison of the two filtered sets of post-fit residuals when plotted on an absolute scale (cf. Fig. \ref{fig:fullspectra})  shows that the high value of post-fit residuals forming the band shaped pattern is dominating in frequencies above \unit[\oldstylenums{20}]{\mhz}.

In order to find their source, we investigated the four \textsc{snr}s of frequencies of the \textsc{kbr} assembly. Since they are the fundamental entity used to compute the \textsc{kbr} system noise, which is dominating in the frequencies above \unit[\oldstylenums{20}]{\mhz} \citep{1999Thomas}. The comparison of the \textsc{snr}s and the high-frequency post-fit residuals show that the bands of high value of residuals are dependent on the variations in the \textsc{k-}band \textsc{snr} of \textsc{grace-b}. The value of post-fit residuals are high in the regions along the orbit where the \textsc{k-}band \textsc{snr} of \textsc{grace-b} drops down to \unit[\oldstylenums{550}]{\oldstylenums{0.1}db-\textsc{h}z}, which is well below the defined mission requirements. Since no other \textsc{snr} shows these patterns (cf. Fig. \ref{fig:2k7Kflag_A}), the only source of errors in the post-fit residuals responsible for these bands is the  degraded signal quality of \textsc{k-}band frequency of \textsc{grace-b}, which is reflected in its \textsc{snr}. As investigated by \citet{harvey2016}, the drops in the \textsc{k-}band \textsc{snr} of \textsc{grace-b} are dependent on the temperature variations observed by one of the thermistors located near microwave assembly. Thus, these band forming patterns of high value of post-fit residuals are due to the temperature effects on the ranging frequencies. As the four phase observations (\textsc{k-} and \textsc{k}a-band of \textsc{grace-a  \& b}) are combined to form the range-rates \citep{atbd-grace}, these errors  propagate to the range-rate observations and, consequently to the range-rate residuals.

Another feature which is also present in the high-frequency range-rate residuals $(\hat{e}_{\text{HP}})$ is the signatures related to the \textsc{s}un intrusions into the \textsc{snr}s (cf. Fig. \ref{fig:intruPfits}(a), (b), (c)). As seen in the Fig. \ref{fig:2k7Kflag_A}, all the valid \textsc{snr}s drop during the \textsc{s}un intrusions into the star cameras. However, one possible source responsible for the \textsc{s}un intrusion dependent errors in the high-frequency residuals is the \textsc{k}-band \textsc{snr} of \textsc{{\gracea}} as the drop in its value during \textsc{s}un intrusions is larger (down to \unit[\oldstylenums{550}]{\oldstylenums{0.1}db-\textsc{h}z})  as compared to the \textsc{snr}s of \textsc{k}a- and  \textsc{k-}band  of  \textsc{grace-a} and \textsc{grace-b} respectively. The differences in the effects on \textsc{snr}s are due to differences in the microwave assemblies used in the two \textsc{grace} spacecrafts (for details see \citet{harvey2016}).
\noindent The amplitude of the residuals is high where the \textsc{k}-band \textsc{snr} of \textsc{{\gracea}} drops in the inner ring structure caused by the \textsc{s}un intrusions as shown in Fig. \ref{fig:intruPfits}(a), (b), (c) as a zoomed-in plot.  However, the signatures of the \textsc{s}un intrusion dependent errors  are not as strong as temperature dependent errors in post-fit range-rate residuals. In Fig. \ref{fig:fullspectra}(a) and (c.), we see that the strength of the intrusion dependent errors in the absolute pre-fit residuals is weaker than the temperature dependent errors. It implies that the range-rate observations are more affected by the temperature effects than by the \textsc{s}un intrusion effects. The amplitude of pre-fit range-rate residuals of August 2008 are comparatively higher than the other months. However, the solution converged with the noise level comparable to other months as can be seen in the post-fit range-rate residuals.
So far, we have shown that the errors in the high frequencies  are largely reflected in the post-fit range-rate residuals. However, it is difficult to say that they are completely absorbed by them without leakage of any part of them in to the estimated parameters. Our approach to observe and to quantify this is to compare the differences between the reduced range-rate observations (pre-fit range-rate residuals defined in Section \ref{section2}) with the post-fit range-rate residuals. The differences should show the amount of the signal mapped on to the parameters estimated (Stokes's coefficients, orbital state parameters, accelerometer scale and bias). Thus,  we analyze their differences in the following section.

						\subsection{Contribution of  high-frequency errors in range-rate observations into the estimated gravity field parameters \label{prefit}}
					In order to investigate whether the investigated high-frequency errors of the range-rate observations are propagated into the estimated parameters, in this section, we analyze the absolute of the differences between the pre-fit and the post-fit residuals.
					Note that, the estimated parameters include unknown initial positions of the orbit determination problem $\mathbf{(r, \dot{r})}$, scale $\mathbf{({S_{x}}, {S_{y}}, {S_{z}})}$ and bias of the accelerometer $\mathbf{({b_{x}}, {b_{y}}, {b_{z}})}$ and Stokes's coefficients $(\mathbf{c_{nm}, s_{nm}})$ (cf. Eqn. \ref{eq:estParam}).

					\begin{align}
					  \text{Estimated parameters} = \left[ \begin{array}{c}
					                \mathbf{c_{nm}, s_{nm}}  \\
					                \mathbf{(r, \dot{r})}_{i} \\
					                \mathbf{({S_{x}}, {S_{y}}, {S_{z}})}_{i} \\
					                \mathbf{({b_{x}}, {b_{y}}, {b_{z}})}_{i}
					                \end{array} \right]
					\label{eq:estParam}
					\end{align}
					where $i \in $\{A, B\}.

\begin{figure}[htbp!]
							\begin{minipage}[c]{0.67\linewidth}
					      \includegraphics[width= 1\linewidth]{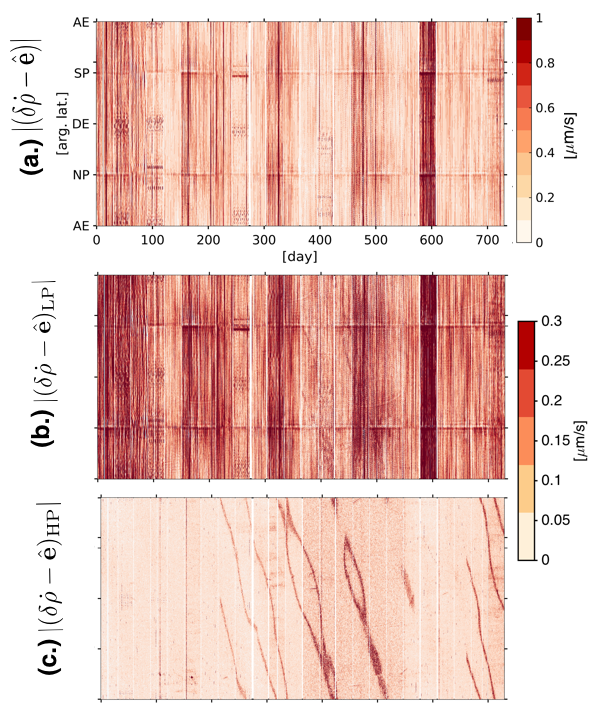}
					  		\end{minipage}\hfill
					  		\begin{minipage}[c]{0.30\linewidth}
					  		\caption{
					  			\textbf{(a.)} shows the differences between the  pre-fit and post-fit residuals, \textbf{(b.)} shows the low-pass filtered part of the differences between the pre-fit and post-fit residuals and \textbf{(c.)} presents the high-pass filtered part of the differences shown in \textbf{(a.)}.	All values are plotted on an absolute scale. Their statistical descriptions are defined in the Table \ref{table:diffDesc}.}
					        \label{fig:diffPfits}
					  		\end{minipage}
						\end{figure}

					The absolute differences between the pre-fit and post-fit residuals should indicate the signal that has been absorbed by the estimated parameters (cf. Eqn. \ref{eq:estParam}). Although the {\kbr} noise is observed in the high-frequency spectrum, we look at the differences between the full signals, their low-frequency ($<$\unit[\oldstylenums{20}]{\mhz}) parts as well as the high-frequency parts ($>$\unit[\oldstylenums{20}]{\mhz}) altogether in Fig. \ref{fig:diffPfits} plotted on an absolute scale.

					The differences between the pre-fit  and post-fit residuals (cf. Fig. \ref{fig:diffPfits} \textbf{(a.)}) show that the contribution of low frequencies into the estimated parameters is significantly higher than the high frequencies. These differences in Fig. \ref{fig:diffPfits} \textbf{(a.)} are highly correlated with the differences of the low-pass filtered parts of the post-fit and pre-fit residuals, i.e. Fig. \ref{fig:diffPfits} \textbf{(b.)}.  The range-rate residuals in the low frequencies ($<$\unit[\oldstylenums{20}]{\mhz}) are  dominated by the attitude errors, accelerometer dependent errors and errors from other unknown sources as discussed in Section \ref{intro}. The analysis of these low-frequency errors in the range-rate residuals is beyond the scope of this paper.
					The differences of the high-frequency filtered set of residuals plotted on the different color scale (cf. Fig. \ref{fig:diffPfits} \textbf{(c.)}) shows the noise that is mapping into the estimated parameters.

					\begin{table}[htbp!]
					  \caption{Statistical description of the differences shown in Fig. \ref{fig:diffPfits}, their high-pass filtered and low-filtered parts shown in the same figure.}
					  \begin{center}
					  \begin{tabular}{p{0.20\textwidth} p{0.20\textwidth} p{0.24\textwidth} p{0.24\textwidth}}
					  \toprule
					  & \small Differences &  \small  	Low-pass filtered				&  \small  High-pass filtered \\
					  \toprule
					  & \textbf{(a.)}  $(\delta \dot{\rho} - \mathbf{\hat{e}})$ &  \textbf{(b.)} $(\delta \dot{\rho} - \mathbf{\hat{e}})_{\text{LP}}$ &  \textbf{(c.)} $(\delta \dot{\rho} - \mathbf{\hat{e}})_{\text{HP}}$ \\
					  \midrule
					  Mean ($\unitfrac{\mu m}{s}$)     & $0.19654$ & $0.19431$ & $0.00384$   \\
					  RMS  ($\unitfrac{\mu m}{s}$)     & $0.662431$ & $0.65859$ & $0.01926$\\
					  Median ($\unitfrac{\mu m}{s}$)   & $0.124630$ & $0.12372$ & $0.00183$ \\
					  \multicolumn{4}{c}{} \\
					  Ratio of mean values &  &$\left|\frac{\text{\textbf{(b.)}}}{\text{\textbf{(a.)}}}\right| \,=\, 0.98865$ & $\left|\frac{\text{\textbf{(c.)}}}{\text{\textbf{(a.)}}}\right| \,=\, 0.01957$ \\
					  Ratio of median values & & $0.99276$ & $0.01471$ \\
					  \midrule
					  \end{tabular}
					  \end{center}
					  \label{table:diffDesc}
					\end{table}
					Ratios in the column \textbf{(c.)} of Table  \ref{table:diffDesc}  explains the amount of high-frequency filtered noise to the total noise mapped into the estimated parameters (cf. Eqn. \ref{eq:estParam}). Similarly, the amount of  low-frequency noise mapped in to the estimated parameters is explained in the ratios of column \textbf{(b.)} of Table \ref{table:diffDesc}.
					The ratios are computed for the mean and median values both. The median is more robust to the outliers whereas mean value is less. Hence, we take the both statistical descriptors into account in order to explain the amount of high-frequency filtered noise mapped in to the estimated parameters. 
					In order to compute the ratios, first, we compute the differences between pre-fit and post-fit residuals. Second, we take the low-pass and high-pass filtered parts of the computed differences. Finally, we compute the mean and median of the differences and their low-pass and high-pass filtered parts. The absolute of the mean and median values are presented in the Table \ref{table:diffDesc} . The ratios are computed from the absolute values computed for each i.e. differences of pre-fit and post-fit residuals, their low-pass and high-pass filtered parts.

					From the ratios explained in Table   \ref{table:diffDesc}, it is clear that the contribution of the low-frequency noise to the estimated parameters is significant as compared to the high-frequency noise. Both, ratios of the mean and median values show that the contribution of high-frequency errors is as small as $\approx$ \unit[\oldstylenums{1}]{\%} whereas the contribution of the low-frequency errors is $\approx$ \unit[\oldstylenums{99}]{\%} in to the estimated parameters.
					However, the contribution of the high-frequency part is reaching up to \unit[\oldstylenums{30}]{\%} of the total error contribution in the months where the temperature dependent non-stationary errors were high (cf. Fig. \ref{fig:diffPfits}\textbf{(c.)}).
					Again, it should be kept in mind that this percentage contribution could be distributed  to any of the parameters estimated (cf. Eqn. \ref{eq:estParam}) during the gravity field processing.

					Since it is clear that the contribution of the high-frequency errors into the estimated parameters is significantly small still, it is worth to model and investigate the impact of these errors on the gravity field solutions in future,  once the full understanding of these errors is established.

					\section{Summary and outlook \label{summary}}
					Our contribution focused on two parts - In the first part we presented an analysis of the \textsc{snr} of the \textsc{k}-band ranging assembly where we present the effects in the \textsc{snr} that were not known before.
					In second part, we have shown that the high \textsc{kbr} system noise which leads to the degraded quality of range observations, is responsible for the noise in high-frequency range-rate residuals.

					First, we presented results of analysis of the \textsc{snr}s of four frequencies on board \textsc{grace}. The analysis of \textsc{snr}s revealed two more systematic effects which were not known. We presented that the \textsc{m}oon intrusions also  affect the quality of the \textsc{snr}s (in Section \ref{moon}). The effect of \textsc{m}oon intrusions  into \textsc{snr}s repeats every \unit[26]{d}. For most of the duration, the drop in the \textsc{snr} values was not below mission requirements but we show that there are periods when the \textsc{snr} drops significantly even below mission requirements during \textsc{m}oon intrusions.  Since the \textsc{kbr} assembly of \textsc{grace}-\textsc{f}ollow \textsc{o}n (\textsc{grace-fo}) will be shielded  to protect it from electromagnetic interference between ranging frequencies  and the \textsc{i}nstrument \textsc{p}rocessing \textsc{u}nit, the identified \textsc{m}oon intrusion effects into the star camera are not expected to influence the ranging frequencies in \textsc{grace-fo} (personal communication, Gerhard L. Kruizinga on 10 Oct. 2016).

					Further, we presented the source of effects in \textsc{snr}s along the equator which were not explained by \citet{harvey2016}. The effects are found to be dependent on the varying currents in the \textsc{mtq}s (in Section \ref{torque}). We have shown that the currents in the \textsc{mtq}s of \textsc{grace-b} are affecting all the three valid \textsc{snr}s, i.e. \textsc{k-} and \textsc{k}a-band frequency of \textsc{grace-a} and \textsc{k-}band frequency of \textsc{grace-b}. The \textsc{snr}s also contain the \textsc{mtq}s dominant frequency  \unit[\oldstylenums{3.3}]{\mhz}.

					\noindent One possible reason could be the electromagnetic interference between the magnetic torquer rod currents and the frequencies of the \textsc{k}-band ranging assembly. However, the hypothesis has to be studied further. The investigations related to the magnetic torquer rod currents induced signals on the \textsc{grace} observations are ongoing in the \textsc{jpl, nasa} (personal communication, Gerhard L. Kruizinga on 12 Oct. 2017).

					Second, we presented an analysis to study the noise present in  high-frequency  range-rate observations in Section \ref{section2}. The quality of the high -frequency range-rate observations is highly affected by the instrument temperature  variations and intrusions   in the star cameras, which is reflected in terms of degraded \textsc{snr} values. Errors due to the temperature variations and the \textsc{s}un intrusions are well reflected in the range-rate residuals. We have shown in Section \ref{prefit} that a significantly small part of the high-frequency errors is  absorbed by the parameters estimated (see Eqn. \ref{eq:estParam} for the list of estimated parameters) during gravity field parameter estimation.

					As we mentioned in Section \ref{torque} that the investigations are still ongoing in \textsc{jpl, nasa} in order to understand such effects, a model needs to be developed after the establishment of their full understanding.
					The model and their full understanding are required to investigate their impact on gravity field and also to mitigate such errors during the pre-processing step in \textsc{grace} gravity field modeling. \newline
					Considering the \textsc{grace-f}ollow \textsc{o}n (\textsc{grace-fo}), it is difficult to predict the nature of errors which would affect the ranging quality before its launch. However,
					this study can be used as a basis to investigate  the errors in the range-rate residuals and to find their sources in early stage of the mission, in order to benefit from the \textsc{grace-fo}.
					An understanding of the errors propagating to the range observations during the initial stage of \textsc{grace-fo} can be helpful in many ways, such as -- finding the possibility to correct them, for example, by satellite maneuvers, and developing the better data processing strategies or noise modeling approaches to mitigate the propagation of these errors in to the gravity field solutions.
					\newpage
					\section*{Acknowledgments}
					 We acknowledge support from the German Research Foundation DFG within SFB 1128 geo-Q to fund this research. We acknowledge Prof. Jakob Flury for the resources and the opportunity he provided us to work on this research topic. We acknowledge the discussions with Gerhard L.Kruizinga, Tamara Bandikova regarding an insight into the sensors data and the ongoing work in this direction.


\bibliographystyle{abbrv}
\end{document}